\documentstyle[11pt]{article}
\def\ATMP#1#2#3{{\it Adv. Theor. Math. Phys.} {\bf #1} {(#2)} {#3}}

\def\NPB#1#2#3{{\it Nucl. Phys.} {\bf B#1} {(#2)} {#3}}
\def\PLB#1#2#3{{\it Phys. Lett.} {\bf B#1} {(#2)} {#3}}
\def\CQG#1#2#3{{\it Class. Quantum Grav.} {\bf #1} {(#2)} {#3}}
\def\MPLA#1#2#3{{\it Mod. Phys. Lett.} {\bf A#1} {(#2)} {#3}}

\def\PRD#1#2#3{{\it Phys. Rev.} {\bf D#1} {(#2)} {#3}}
\def\PR#1#2#3{{\it Phys. Rep.} {\bf #1} {(#2)} {#3}}

\begin{document}
{}~\hfill FIAN/TD/23-99

{}~\hfill hep-th/9908114

\vspace{3cm}

\begin{center}

{\Large\bf
Light cone gauge formulation of $IIB$ supergravity in

\medskip
$AdS_5 \times S^5$ background and AdS/CFT correspondence}

\vspace{2cm}
R.R. Metsaev\footnote{e-mail: metsaev@lpi.ru}

\vspace{1cm}
{\it Department of Theoretical Physics, P.N. Lebedev Physical
Institute, Leninsky prospect 53, 117924, Moscow, Russia}

\vspace{3cm}
{\bf Abstract}
\end{center}

\noindent
Light cone gauge manifestly  supersymmetric formulations of type
 IIB  10-dimensional supergravity
in $AdS_5 \times S^5$ background and  related  boundary conformal
field theory representations are developed.
 A precise correspondence
between the bulk fields of IIB supergravity and the boundary operators
is established. The formulations are given entirely in terms of light cone
scalar superfields, allowing  us to treat all component fields on
an equal footing.

\vspace{3cm}

\begin{center}
Published in \PLB{468}{1999}{65-75}
\end{center}

\newpage

{\bf Introduction}.
A long-term motivation  for  our investigation comes from the following
potentially important application. Inspired by  the conjectured duality
between the string theory and ${\cal N}=4$, $4d$ SYM theory
\cite{mal} the Green-Schwarz formulation of strings propagating
in $AdS_5\times S^5$ was suggested in \cite{mt1} (for  further
developments see \cite{krr}-\cite{k1}).  Despite considerable efforts
these strings have not yet been quantized (some related interesting
discussions are in \cite{Dolan:1999pi}).
Alternative approaches can be found in \cite{alter}. As is
well known, quantization of GS superstrings propagating in flat space is
straightforward only in the light cone gauge.
It is the light cone gauge
that removes unphysical degrees of freedom explicitly and reduces the
action to quadratic form in string coordinates.  The light cone gauge in
string theory implies the corresponding light cone formulation for target
space fields. The string theories are approximated at low energies  by
supergravity theories. This suggests that we should first study a light
cone gauge formulation of supergravity theory in AdS spacetime.
Understanding a light cone description of type $IIB$ supergravity in
$AdS_5\times S^5$ background might help to solve problems of strings in
AdS spacetime.

Keeping in mind extremely important applications to string theory, in this
paper we develop the light cone gauge formulation of $IIB$ supergravity
in $AdS_5\times S^5$ and the associated  boundary CFT and apply our results to
the study of AdS/CFT correspondence at the level of state/operator matching.
Our method is conceptually very close to the one used
in \cite{GS5} (see also \cite{GSB}) to find the light cone form of $IIB$
supergravity in  flat space and is based essentially on a light cone gauge
description of field dynamics developed recently in \cite{metlc}.
A discussion of $IIB$ supergravity at the level of
 gauge invariant equations of motion and
actions can be found in \cite{covequmot} and \cite{covact} respectively.
As is well known in the case of the massless fields, investigation of
AdS/CFT correspondence requires the analysis of some subtleties related to the
fact that transformations of  massless bulk fields are defined up to local
gauge transformations. These complications are absent in the light cone
formulation because here we deal only  with the physical fields, and   this allows
us to demonstrate the AdS/CFT correspondence in a rather straightforward way.

\medskip
{\bf Light cone form of $psu(2,2|4)$ superalgebra and notation}.
We use the  following  parametrization of $AdS_5\times S^5$ space
$$
ds^2=\frac{1}{z^2}(-dt^2+dx_1^2+dx_2^2+dz^2+ dx_4^2)
+\frac{1}{4}dy_{ij}dy_{ij}^*
\,,
\qquad
z>0\,.
$$
Here and below we set the radii of  both  $AdS_5$ and $S^5$ equal to unity.
The boundary at spatial infinity corresponds to $z=0$. The
$S^5$ coordinates $y^{ij}$ are subject to the constraints
$$
y^{ik}y_{kj}=\delta_j^i\,,
\qquad
y_{ij}=\frac{1}{2}\epsilon_{ijkn}y^{kn}\,,
\qquad
y_{ij}^*=-y^{ij}\,,
\qquad
i,j,k,n=1,2,3,4
$$
where $\epsilon^{ijkn}=\pm 1$ is the Levi-Civita  tensor of $su(4)$.
The coordinates $y_{ij}$ are related to the  standard $so(6)$
cartesian coordinates $y^M$, $M=1,\ldots,6$, which satisfy the constraint
$y^My^M=1$ through the formula $y_{ij}=\rho_{ij}^M y^M$,
where $\rho_{ij}^M$ are the Clebsh Gordan coefficients of $su(4)$
algebra \cite{GS6}. We use the coordinates $y^{ij}$ instead of $y^M$ as
this allows us to avoid  using  various cumbersome gamma matrix
identities. To develop light cone formulation we introduce light cone
variables $x^\pm \equiv(x^4  \pm x^0)/\sqrt{2}$, $x\equiv(x^1+{\rm
i}x^2)/\sqrt{2}$, $\bar{x}\equiv x^*$, where $x^0\equiv t$. In the
following we treat $x^{+}$ as evolution parameter.

Now let us discuss the form of the algebra of isometry transformations
of $AdS_5\times S^5$ superspace, that is $psu(2,2|4)$,  which we are going to
use.  The even part of this algebra is the algebra $so(4,2)$ which is the
isometry algebra of $AdS_5$ and the algebra $so(6)$ which is the  isometry
algebra of $S^5$. The odd part of the  algebra consists of 32 supercharges
which are responsible for 32 Killing spinors in $AdS_5\times S^5$ space.
We prefer to use the form of $so(4,2)$ algebra provided by nomenclature of
conformal algebra. In this notation we have, as usual, translations
$P^a$, conformal boosts $K^a$, dilatation $D$ and Lorentz rotations
$J^{ab}$ which satisfy the commutation relations
\begin{eqnarray}
&&{}\!\!\!\!\!\!\!\!
[D,P^a]=-P^a,
\quad
[D,K^a]=K^a,
\quad
[P^a,J^{bc}]=\eta^{ab}P^c-\eta^{ac}P^b,
\quad
[K^a,J^{bc}]=\eta^{ab}K^c-\eta^{ac}K^b,
\nonumber\\
{}
\label{comrel1}
%&&\hspace{3cm}
&&
[P^a,K^b]=\eta^{ab}D-J^{ab},
\qquad
[J^{ab},J^{cd}]=\eta^{bc}J^{ad}+ 3\hbox{ terms}\,,
\quad
a,b,c,d=0,1,2,4
\end{eqnarray}
where $\eta^{ab}=(-,+,+,+)$. Throughout this paper instead of $so(6)$
algebra notation we prefer to use the  notation of $su(4)$ algebra.
Commutation relation of $su(4)$ algebra are
$$
[J^i{}_j,J^k{}_n]=\delta^k_jJ^i{}_n
-\delta^i_n J^k{}_j\,.
$$
In the light cone basis we have: $J^{\pm x}$, $J^{\pm\bar x}$, $J^{+-}$,
$J^{x\bar x}$, $P^\pm$, $P^x$, $P^{\bar x}$, $K^\pm$, $K^x$, $K^{\bar x}$.
We simplify notation as follows
$P\equiv P^x$, $\bar{P}=P^{\bar x}$, $K\equiv K^x$, $\bar{K}=K^{\bar x}$.
The light cone form of $so(4,2)$ algebra commutation relations can be
obtained from (\ref{comrel1}) with the light cone metric having the
following non vanishing elements $\eta^{+-}=\eta^{-+}=1$,
$\eta^{x\bar{x}}=\eta^{\bar{x}x}=1$.

To describe the odd part of $psu(2,2|4)$ superalgebra we introduce
32 supercharges $Q^{\pm i}$, $Q_i^\pm$, $S^{\pm i}$, $S_i^\pm$ which
possess $D$, $J^{+-}$ and $J^{x\bar x}$ charges.
The commutation relations of supercharges with dilatation $D$
\begin{equation}
[D,Q^{\pm i}]=-\frac{1}{2}Q^{\pm i}\,,
\quad
[D,Q^\pm_i]=-\frac{1}{2}Q^\pm_i\,,
\quad
[D,S^{\pm i}]=\frac{1}{2}S^{\pm i}\,,
\quad
[D,S^\pm_i]=\frac{1}{2}S^\pm_i\,,
\end{equation}
tell us that $Q$ are usual supercharges
of Poincar\'e subsuperalgebra, while $S$ are conformal supercharges.
The supercharges with superscript $+$ ($-$) have positive (negative)
$J^{+-}$ charge
$$
[J^{+-},Q^{\pm i}]
=\pm\frac{1}{2}Q^{\pm i}\,,
\quad
[J^{+-},Q^\pm_i]
=\pm\frac{1}{2}Q^\pm_i\,,
\quad
[J^{+-},S^{\pm i}]
=\pm\frac{1}{2}S^{\pm i}\,,
\quad
[J^{+-},S^\pm_i]
=\pm\frac{1}{2}S^\pm_i\,.
$$
The $J^{x\bar x}$ charges are fixed by the commutation relations
\begin{equation}
[J^{x\bar{x}},Q^{\pm i}]=\pm\frac{1}{2}Q^{\pm i}\,,
\quad
[J^{x\bar{x}},Q^\pm_i]=\mp\frac{1}{2}Q^\pm_i\,,
\quad
[J^{x\bar{x}},S^\pm_i]=\pm\frac{1}{2}S^\pm_i\,,
\quad
[J^{x\bar{x}},S^{\pm i}]=\mp\frac{1}{2}S^{\pm i}\,.
\end{equation}
Transformation properties of supercharges with respect to $su(4)$
algebra are given by
$$
[Q^\pm_i,J^j{}_k]=\delta_i^jQ^\pm_k-\frac{1}{4}\delta^j_kQ^\pm_i\,,
\qquad
[Q^{\pm i},J^j{}_k]=-\delta^i_k Q^{\pm j}
+\frac{1}{4}\delta^j_kQ^{\pm i}
$$
and the same for supercharges $S$.
Anticommutation relations between supercharges are
$$
\{Q^{\pm i},Q_j^\pm\}=\pm P^\pm\delta^i_j\,,
\quad
\{Q^{+i},Q_j^-\}=P\delta^i_j\,,
\quad
\{S^{\pm i},S_j^\pm\}=\pm K^\pm\delta^i_j\,,
\quad
\{S^{-i},S_j^+\}=K\delta^i_j\,,
$$
$$
\{Q^{+i},S^+_j\}=-J^{+x}\delta^i_j\,,
\quad
\{Q^{-i},S^-_j\}=-J^{-\bar{x}}\delta^i_j\,,
$$
$$
\{Q^{\pm i},S^\mp_j\}=\frac{1}{2}(J^{+-}+J^{x\bar{x}}\mp D)\delta_j^i
\mp J^i{}_j\,.
$$
Remaining commutation relations between supercharges and even part of
superalgebra take the following form
$$
[Q^{-i},J^{+x}]=-Q^{+i}\,,
\quad
[S^{-i},J^{+\bar{x}}]=-S^{+i}\,,
\quad
[Q^{+i},J^{-\bar{x}}]=Q^{-i}\,,
\quad
[S^{+i},J^{-x}]=S^{-i}\,,
$$
$$
[S^\mp_i,P^\pm]=Q^\pm_i\,,
\quad
[S^-_i,P]=Q^-_i\,,
\quad
[S^+_i,\bar{P}]=-Q^+_i\,,
$$
$$
[Q^{\mp i},K^\pm]=S^{\pm i}\,,
\quad
[Q^{-i},K]=S^{-i}\,,
\quad
[Q^{+i},\bar{K}]=-S^{+i}\,.
$$
The above generators are subject to  the following hermitean
 conjugation  conditions
$$
P^{\pm\,\dagger}=P^\pm,
\quad
P^\dagger=\bar{P},
\quad
(K^\pm)^\dagger=K^\pm,
\quad
K^\dagger=\bar{K},
\quad
(Q^{\pm i})^\dagger=Q^{\pm}_i,
\quad
(S^{\pm i})^\dagger=S^{\pm}_i\,,
$$
$$
(J^{\pm x})^\dagger=-J^{\pm \bar{x}}\,,
\quad
(J^{+-})^\dagger=-J^{+-}\,,
\quad
(J^{x\bar{x}})^\dagger=J^{x\bar{x}}\,,
\quad
D^\dagger=-D\,,
\quad
J^i{}_j^\dagger=J^j{}_i\,.
$$
All the remaining nontrivial (anti)commutation relations of $psu(2,2|4)$
superalgebra could be obtained by using these hermitean conjugation rules
and (anti)commutation relations given above.

\medskip
{\bf Light cone gauge formulation of $IIB$ supergravity in $AdS_5\times
S^5$ background}.
Our next step is to find  a realization of $psu(2,2|4)$
superalgebra on the space of IIB supergravity fields propagating in
$AdS_5\times S^5$.  To do that we use light cone superspace
formalism. First,  we introduce light cone superspace which by definition
is based on position $AdS_5\times S^5$ coordinates $x^\pm$, $x$,
$\bar{x}$, $z$, $y^{ij}$ and Grassmann position coordinates $\theta^i$ and
$\chi^i$ which transform in fundamental ${\bf 4}$ irreps of $su(4)$
algebra. Second,  on this light cone superspace we introduce scalar
superfield $\Phi(x^\pm,x,\bar{x},z,y^{ij},\theta^i,\chi^i)$. In the  remainder
of this paper we find it convenient to Fourier transform to momentum space
for all coordinates except for the radial $z$ and $S^5$ coordinates
$y^{ij}$.  This implies using $p^-$, $p^+$, $\bar{p}$, $p$, $\lambda_i$,
$\tau_i$ instead of $x^+$, $x^-$, $x$, $\bar{x}$, $\theta^i$, $\chi^i$
respectively. The $\lambda_i$ and $\tau_i$ transform in
$\bar{{\bf 4}}$ irreps of $su(4)$. Thus we consider the superfield
$\Phi(p^\pm,p,\bar{p},z,y^{ij},\lambda_i,\tau_i)$ with the following
expansion in powers of Grassmann momenta $\lambda_i$ and $\tau_i$
\begin{eqnarray}
\Phi
&=&p^{+2}\phi
+p^+\Bigl(\lambda_i\psi_1^i+\tau_i\psi_2^i\Bigr)
\nonumber\\
&+&
p^+\Bigl(\lambda_i\lambda_j\phi_1^{ij}
+\lambda_i\tau_j\phi_2^{ij}+\tau_i\tau_j\phi_3^{ij}\Bigr)
\nonumber\\
&+&
(\epsilon\lambda^3)^i\psi_{3i}+\lambda_i\lambda_j\tau_k\psi_1^{ijk}
+\lambda_i\tau_j\tau_k\psi_2^{ijk}
+(\epsilon\tau^3)^i\psi_{4i}
\nonumber\\
&+&
(\epsilon\lambda^4)\phi_4
+(\epsilon\lambda^3)^i\tau_j\phi^j{}_i
+\lambda_i\lambda_j\tau_k\tau_n\phi^{ijkn}
+\lambda_i(\epsilon\tau^3)^j\phi^j{}^*_i
+(\epsilon\tau^4)\phi_4^*
\nonumber\\
&+&
\frac{1}{p^+}\Bigl(
-\lambda_i(\epsilon\tau^4)\psi_{3i}^*
+(\epsilon\lambda^2)^{ij}(\epsilon\tau^3)^k\psi_1^{ijk*}
+(\epsilon\lambda^3)^i(\epsilon\tau^2)^{jk}\psi_2^{ijk*}
-(\epsilon\lambda^4)\tau_i\psi_{4i}^*
\Bigr)
\nonumber\\
&+&
\frac{1}{p^+}\Bigl(
-(\epsilon\lambda^2)^{ij}(\epsilon\tau^4)\phi_1^{ij*}
+(\epsilon\lambda^3)^i(\epsilon\tau^3)^j\phi_2^{ij*}
-(\epsilon\lambda^4)(\epsilon\tau^2)^{ij}\phi_3^{ij*}
\Bigr)
\nonumber\\
\label{supfield}
&-&
\frac{1}{p^{+2}}\Bigl(
(\epsilon\lambda^3)^i(\epsilon\tau^4)\psi_1^{i*}
+(\epsilon\lambda^4)(\epsilon\tau^3)^i\psi_2^{i*}
\Bigr)
+\frac{1}{p^{+2}}(\epsilon\lambda^4)(\epsilon\tau^4)\phi^*
\end{eqnarray}
where
$$
(\epsilon \lambda^4)
\equiv \frac{1}{4!}\epsilon^{ijkn}
\lambda_i\lambda_j\lambda_k\lambda_n\,,
\qquad
(\epsilon \lambda^3)^i
\equiv \frac{1}{3!}\epsilon^{ijkn}
\lambda_j\lambda_k\lambda_n\,,
\qquad
(\epsilon \lambda^2)^{ij}
\equiv \frac{1}{2!}\epsilon^{ijkn}
\lambda_k\lambda_n\,,
$$
and the same notation is adapted for $\tau_i$. The field $\phi^{ijkn}$
satisfies the constraint
$$
\phi^{ijkn}=\frac{1}{4}\epsilon^{iji_1j_1}\epsilon^{knk_1n_1}
\phi^{i_1j_1k_1n_1*}\,.
$$
In (\ref{supfield}) the $\phi$ is a complex scalar field,
the $\psi_1^i$, $\psi_2^i$ are two spin one half fields,
the fields $\phi_{1,2,3}^{ij}$ describe two Kalb Ramond fields,
the $\psi^{ijk}_{1,2}$, $\psi_{3,4}^i$ are two gravitinos, while $\phi_4$,
$\phi^i{}_j$, $\phi^{ijkn}$ describe graviton and self dual 4 form field.
The reality constraint in terms of the superfield $\Phi$ takes the form
$$
\Phi(-p, z, y,\lambda,\tau)
=p^{+4}\int d^4\lambda^\dagger d^4\tau^\dagger
e^{(\lambda_i\lambda_i^\dagger+\tau_i\tau_i^\dagger)/p^+}
(\Phi(p,z, y,\lambda,\tau))^\dagger\,.
$$
In  the light cone formalism the $psu(2,2|4)$ superalgebra has
 the generators
\begin{equation}\label{kingen}
P^+,\,P,\,\bar{P},\, J^{+x},\,J^{+\bar x},\, K^+,\, K,\,{\bar K},\,
Q^{+i},\,Q^+_i,\,S^{+i},\,S^+_i,\,
D,\,J^{+-},\,J^{x\bar x}\,,
\end{equation}
which we refer to as kinematical generators,  and
\begin{equation}\label{dyngen}
P^-,\, J^{-x}\,, J^{-\bar x},\,K^-\,, Q^{-i},\, Q^-_i,\, S^{-i},\,S^-_i
\end{equation}
which we refer to as dynamical generators. The kinematical generators
have positive or zero $J^{+-}$ charges, while dynamical generators have
negative $J^{+-}$ charges. For $x^+=0$ the kinematical generators are
quadratic in the physical field $\Phi$, while the dynamical
generators receive corrections in interaction theory.  In this paper we
deal with free fields. At a quadratic level both kinematical and
dynamical generators have the following representation in
terms of the physical light cone
superfield

$$
\hat{G}=\int dp^+d^2p dz dS^5d^4\lambda d^4\tau\, p^+
\Phi(-p,z,y,-\lambda,-\tau) G\Phi(p,z,y,\lambda,\tau)\,,
$$
where $G$ are the differential operators acting on $\Phi$.
Thus we should find representation of $psu(2,2|4)$ in terms of
differential operators acting on light cone scalar superfield $\Phi$.
To simplify expressions let us write down the generators
for $x^+=0$. The  kinematical generators are then given by
\begin{equation}\label{kin1}
P=p\,,
\qquad
\bar{P}=\bar{p}\,,
\qquad
P^+=p^+\,,
\qquad
J^{+x}=\partial_pp^+\,,
\qquad
J^{+\bar{x}}=\bar{\partial}_pp^+\,,
\end{equation}
\begin{equation}\label{kin2}
K^+=\frac{1}{2}(z^2-2\partial_p\bar{\partial}_p)p^+\,,
\qquad
K=K_0+\frac{1}{2}\partial_p+\theta^iS_i^+\,,
\end{equation}
\begin{equation}\label{kin3}
Q^{+i}=p^+\theta^i\,,
\qquad
Q_i^+=\lambda_i\,,
\qquad
S^+_i=\frac{1}{\sqrt{2}}z\tau_i-\lambda_i\partial_p\,,
\qquad
S^{+i}=\frac{1}{\sqrt{2}}zp^+\chi^i+p^+\theta^i\bar{\partial}_p\,,
\end{equation}
\begin{eqnarray}
&&
J^{+-}=\partial_p^-p^+-\frac{1}{2}\theta\lambda
-\frac{1}{2}\chi\tau+2\,,
\hskip10em
J^{x\bar{x}}=p\bar{\partial}_p-\bar{p}\partial_p
+\frac{1}{2}\theta\lambda-\frac{1}{2}\chi\tau,
\nonumber\\
\label{kin4}
&&
D=-\partial_p^-p^+
-\partial_p\bar{p}-\bar{\partial}_pp+z\partial_z
+\frac{1}{2}\theta\lambda+\frac{1}{2}\chi\tau
-\frac{1}{2}\,,
\qquad
J^i{}_j=l^i{}_j+M^i{}_j\,,
\end{eqnarray}
where we use the  notation
$$
K_0\equiv \frac{1}{2}(z^2-2\partial_p\bar{\partial}_p)p
-\partial_p(-\partial_p^- p^+ -\partial_p\bar{p}
-\bar{\partial}_pp+z\partial_z)\,,
$$
\begin{equation}\label{lm}
l^i{}_j=\frac{1}{2}y^{ik}\nabla_{kj}\,,
\quad
M^i{}_j\equiv \theta^i\lambda_j
+\chi^i\tau_j-\frac{1}{4}\delta^i_j(\theta\lambda+\chi\tau),
\quad
\nabla_{ij}\equiv \rho_{ij}^M(\delta^{MN}-y^My^N)\partial_{y^N}\,,
\end{equation}
$$
\partial_p^\pm\equiv \partial/\partial p^\mp,
\quad
\partial_p\equiv \partial/\partial \bar{p},
\quad
\bar{\partial}_p\equiv \partial/\partial p\,,
\quad
\theta\lambda\equiv \theta^i\lambda_i\,,
\quad
\chi\tau\equiv \chi^i\tau_i\,.
$$
The orbital part of $su(4)$ angular momentum $l^i{}_j$
satisfies the following important relations
$$
l^i{}_ml^m{}_j=\frac{1}{4}l^m{}_nl^n{}_m\delta_j^i+2l^i{}_j\,,
\qquad
[l^i{}_j,l^k{}_n]=\delta^k_jl^i{}_n
-\delta^i_n l^k{}_j\,,
$$
frequently used in this paper. Dynamical generators are given by
\begin{equation}\label{adspm}
P^-=-\frac{p\bar{p}}{p^+}+\frac{\partial^2_z}{2p^+}
-\frac{1}{2z^2p^+}A\,,
\end{equation}
\begin{equation}\label{jmx}
J^{-x}\!=-\partial_p^-p+\partial_pP^-
+\theta^iQ_i^- -\frac{2p}{p^+},
\quad
J^{-\bar x}\!=-\partial_p^-\bar{p} +
\bar{\partial}_pP^-
-Q^{-i}\frac{\lambda_i}{p^+}
+(\theta\lambda+\chi\tau-2)\frac{\bar{p}}{p^+}\,,
\end{equation}
\begin{equation}\label{adsqm}
Q^{-i}=\bar{p}\theta^i+\frac{1}{\sqrt{2}}\Bigl(\chi^i\partial_z
+\frac{1}{2z}[\chi^i,A]\Bigr)\,,
\qquad
Q^-_i=\frac{p}{p^+}\lambda_i
-\frac{1}{\sqrt{2}p^+}\Bigl(
\tau_i\partial_z+\frac{1}{2z}[\tau_i,A]\Bigr)\,,
\end{equation}
where
\begin{equation}\label{ax}
A\equiv X-\frac{1}{4}\,,
\quad
X\equiv l^i{}_j^2+4\tau l \chi +(\chi\tau-2)^2,
\quad
l^i{}_j^2\equiv l^i{}_jl^j{}_i\,,
\quad
\tau l\chi \equiv\tau_i l^i{}_j\chi^i.
\end{equation}
For $\lambda_i$, $\tau_i$ and $\tau^i$, $\chi^i$ we adopt the following
anticommutation and hermitean conjugation rules
$$
\{\theta^i,\lambda_j\}=\delta^i_j\,,
\quad
\{\chi^i,\tau_j\}=\delta^i_j\,,
\quad
\lambda_i^\dagger=p^+\theta^i, \quad \theta^{i
\dagger}=\frac{1}{p^+}\lambda_i\,, \quad \tau_i^\dagger=p^+\chi^i, \quad
\chi^{i \dagger}=\frac{1}{p^+}\tau_i\,.
$$
Remaining generators $\bar{K}$, $S^{-i}$, $S^-_i$, $K^-$ are
obtainable from the above generators  via commutation relations
of $psu(2,2|4)$ superalgebra. Because these expressions are
not illuminating we do not present them here.

 Following \cite{metlc}
we shall call  the
operator $A$  the $AdS$ mass operator. Few comments are in order.
(i)  The operator $A$ is equal to zero only for massless representations
which can be realized as irreducible representations of conformal algebra
\cite{metsit3},\cite{metlc} which for the case of $AdS_5$
space is the $so(5,2)$ algebra. Below we shall demonstrate that
operator $X$ (\ref{ax}) has eigenvalues equal to squared integers
in the whole spectrum of compactification of $IIB$ supergravity on $AdS_5$.
This implies that operator $A$ (\ref{ax}) is never equal to zero. From
this we conclude that the  scalar fields \cite{krn} as well as all remaining
fields of compactification of $IIB$ supergravity do not satisfy conformally
invariant equations of motion.
(ii) Generators involve the terms linear and quadratic in Grassmann
variables $\theta^i$, $\lambda_i$ and terms up to fourth power in
$\chi^i$ and $\tau_i$. The coordinate $\theta^i$ (or $\lambda_i$)
constitutes odd part of light cone superspace appropriate to
superfield description of light cone gauge $N=4$, $4d$ SYM theory. The
terms of the third and fourth powers in $\chi^i$ and $\tau_i$ are
expressible in terms of the operator $A$ and its commutators with $\chi^i$ and
$\tau_i$.

Recall that the above representation was  given for $x^+=0$.
To study AdS/CFT correspondence we shall need the generators for
arbitrary $x^+\equiv {\rm i}\partial_p^+$ which are given by
$$
G_{x^+}=e^{-\partial_p^+P^-}G_{x^+=0}e^{\partial_p^+P^-},
\qquad
G_{x^+}=G_{x^+=0}
-\partial_p^+[P^-,G_{x^+=0}]
+\frac{1}{2}\partial_p^{+ 2}[P^-,[P^-,G_{x^+=0}]]\,.
$$
Using these relations we derive the complete
expression for conformal supergenerators
$$
S^+_i=S_i^+|_{x^+=0}+\partial_p^+Q^-_i\,,
\qquad
S^{+i}=S^{+i}|_{x^+=0}-\partial_p^+Q^{-i}\,.
$$
Below we shall need the complete expression for the following generators
\begin{equation}\label{adsjpcom}
J^{+x}=\partial_p p^+ - \partial_p^+ p\,,
\quad
J^{+\bar x}=\bar{\partial}_p p^+ - \partial_p^+ \bar{p}\,,
\quad
J^{+-}=J^{+-}|_{x^+=0}-\partial_p^+P^-,
\end{equation}
\begin{equation}\label{adskmcom}
K^+=\frac{1}{2}(z^2-2\partial_p\bar{\partial}_p)p^+
-\partial_p^+(-\partial_p^+P^-
-\partial_p\bar{p}-\bar{\partial}_pp
+z\partial_z+\frac{3}{2})\,,
\qquad
D=D|_{x^+=0}-\partial_p^+P^-.
\end{equation}
Making  use of the expression for $P^-$ (\ref{adspm}) we can immediatelly
write down the light cone gauge action
$$
S_{l.c.}
=\int d^4pdz dS^5d^4\lambda d^4\tau\,
p^+\Phi(-p,z,y,-\lambda,-\tau)
(-p^-+P^-)\Phi(p,z,y,\lambda,\tau)\,.
$$
Since the action is invariant with respect to the  symmetries generated by
$psu(2,2|4)$ superalgebra,  the formalism we discuss is  sometimes referred
to as an off shell light cone formulation \cite{GSB}.

In what follows we shall exploit the following above mentioned important
property of the operator $X$ (\ref{ax}) -- that its
eigenvalues are squared integers. Let us demonstrate this
important fact. First,  we expand the scalar superfield (\ref{supfield})
in $S^5$ coordinates $y^{ij}$ and Grassmann momentum $\tau_i$
\begin{equation}
\Phi=\sum_{l=0}^\infty \Phi_l\,,
\qquad\qquad
\Phi_l=\sum_{a=0}^4\Phi_{l,a}\,,
\end{equation}
where `spherical' harmonic superfields $\Phi_{l,a}$ satisfy the constraints
\begin{equation}\label{deccon}
l^i{}^2_j\Phi_{l,a}=l(l+4)\Phi_{l,a}\,,
\qquad\qquad
\tau\chi \Phi_{l,a}=a\Phi_{l,a}\,.
\end{equation}
The first constraint tells us that $\Phi_{l,a}$ is an eigenvalue vector
of square of $su(4)$ orbital momentum $l^i{}_j$, while the second
constraint tells us that $\Phi_{l,a}$ is a monomial of degree $a$ in
$\tau_i$.  Note that superfield $\Phi_0$ consists of fields of $N=8$, $5d$
AdS supergravity, while the superfields $\Phi_{l>0}$ are responsible for
Kaluza-Klein modes.  Second,  we evaluate eigenvalues of $X$ for each
superfield $\Phi_{l,a}$ in turn. For the case of $\Phi_{l,0}$,
$\Phi_{l,4}$ the eigenvalues of $X$ are easily found to be
\begin{equation}\label{xl0l4}
X\Phi_{l,0}=(l+2)^2\Phi_{l,0}\,,
\qquad
X\Phi_{l,4}=(l+2)^2\Phi_{l,4}\,.
\end{equation}
In deriving the second relation one needs to use the
relation like
\begin{equation}\label{somrel}
\tau_i(\tau l)_j(\tau l)_k\Phi_l^{ijk}
=-\frac{1}{12}l(l+4)\tau_i\tau_j\tau_k\Phi_l^{ijk}\,,
\qquad
(\tau l)_i\equiv \tau_jl^j{}_i\,,
\end{equation}
where $\Phi_l^{ijk}$ is totally antisymmetric in $i,j,k$.
Next,  we analyse the spectrum of $X$ in $\Phi_{l,1}$
and $\Phi_{l,3}$.  In contrast to (\ref{xl0l4}) it turns out that these
superfields themselves do not diagonalize the operator $X$. Now we
decompose the superfields $\Phi_{l,1}$ and $\Phi_{l,3}$ as follows $$
\Phi_{l,1}=\Phi_{l,1}^{(1)}+\Phi_{l,1}^{(2)}\,,
\qquad
\Phi_{l,3}=\Phi_{l,3}^{(1)}+\Phi_{l,3}^{(2)}\,,
$$
where
$$
\Phi_{l,1}^{(1)}
\equiv (\tau_i-\frac{2}{l+4}(\tau l)_i)\Phi_{l,1}^i\,;
\qquad
\Phi_{l,1}^{(2)}
\equiv(\tau_i+\frac{2}{l}(\tau l)_i)\Phi_{l,1}^i,
\quad l>0\,;
$$
$$
\Phi_{l,3}^{(1)}
=(\tau_i\tau_j\tau_k-\frac{6}{l+4}\tau_i\tau_j(\tau l)_k)
\Phi^{ijk}_{l,3}\,;
\qquad
\Phi_{l,3}^{(2)}
=(\tau_i\tau_j\tau_k+\frac{6}{l}\tau_i\tau_j(\tau l)_k)
\Phi^{ijk}_{l,3}\,,
\quad
l>0\,;
$$
and superfields $\Phi_{l,1}^i$, $\Phi_{l,3}^{ijk}$ do not depend on
$\tau_i$. The $\Phi_{l,3}^{ijk}$ is totally antisymmetric in $i,j,k$.
Straightforward calculation gives the following eigenvalues of
operator $X$ $$ X\Phi_{l,1}^{(1)}=(l+1)^2\Phi_{l,1}^{(1)}, \quad
X\Phi_{l,1}^{(2)}=(l+3)^2\Phi_{l,1}^{(2)},
\quad
X\Phi_{l,3}^{(1)}=(l+1)^2\Phi_{l,3}^{(1)},
\quad
X\Phi_{l,3}^{(2)}=(l+3)^2\Phi_{l,3}^{(2)}.
$$
Finally,  we consider the most challenging case of the
 superfield $\Phi_{l,2}$.
It turns out that this superfield is decomposed into three superfields
which are eigenvectors of operator $X$
$$
\Phi_{l,2}=\Phi_{l,2}^{(1)}+\Phi_{l,2}^{(2)}+\Phi_{l,2}^{(3)}\,,
$$
\begin{eqnarray*}
&&
\Phi_{l,2}^{(1)}
\equiv\Bigl(\tau_i\tau_j-\frac{4(l+3)}{(l+2)(l+4)}\tau_i(\tau l)_j
+\frac{4}{(l+2)(l+4)}(\tau l)_i(\tau l)_j\Bigr)\Phi^{ij}_{l,2}\,;
\\
&&
\Phi_{l,2}^{(2)}
\equiv\Bigl(\tau_i\tau_j+\frac{8}{l(l+4)}\tau_i(\tau l)_j
-\frac{4}{l(l+4)}(\tau l)_i(\tau l)_j\Bigr)\Phi^{ij}_{l,2}\,,
\qquad
l>0\,;
\\
&&
\Phi_{l,2}^{(3)}
\equiv\Bigl(\tau_i\tau_j+\frac{4(l+1)}{l(l+2)}\tau_i(\tau l)_j
+\frac{4}{l(l+2)}(\tau l)_i(\tau l)_j\Bigr)\Phi^{ij}_{l,2}\,,
\qquad
l>0\,;
\end{eqnarray*}
where the superfield $\Phi_{l,2}^{ij}$ does not depend on $\tau_i$ and is
totally antisymmetric in $i,j$. Relatively straightforward calculation
gives the following eigenvalues of the operator $X$
$$
X\Phi_{l,2}^{(1)}=l^2\Phi_{l,2}^{(1)}\,,
\qquad
X\Phi_{l,2}^{(2)}=(l+2)^2\Phi_{l,2}^{(2)}\,,
\qquad
X\Phi_{l,2}^{(3)}=(l+4)^2\Phi_{l,2}^{(3)}\,.
$$
Thus we have demonstrated that the operator $X$ in whole space of
superfield $\Phi$ take values which are squares of integers.
This implies  that the operator $\kappa\equiv\sqrt{X}$ is well defined and
possesses integer eigenvalues which are chosen to be positive
in what follows.

\medskip
{\bf Light cone form of boundary CFT}.
The next primary goal of this work is to find a light cone
gauge realization of $psu(2,2|4)$ superalgebra at the boundary of
$AdS_5\times S^5$ which is $M^{3,1}\times S^5$, where $M^{3,1}$ is a (3+1)
Minkowski space time (for a review of CFT see, for instance, \cite{frpal}).
At this boundary the superalgebra acts as the  algebra
of conformal transformations. Now we have to realize this superalgebra on the
space of conformal operators. To this end we introduce boundary
light cone superspace which is based on momentum  variables  $p^\pm$, $p$,
$\bar{p}$, position $S^5$ coordinate $y^{ij}$ and the Grassmann
momentum  variables $\lambda_i$, $\tau_i$. On this light cone
superspace we define a superfield ${\cal O}_{loc}$ which is collection
of CFT operators with canonical dimensions (currents)
and superfield $\tilde{{\cal O}}_{loc}$ which is a collection of shadow
operators (sources). These superfields have a similar expansion in
$\lambda_i$ and $\tau_i$ as the superfield $\Phi$ (\ref{supfield}) does.
In general the representation of $psu(2,2|4)$ algebra  in  ${\cal O}_{loc}$
differs from the one in $\tilde{{\cal O}}_{loc}$. It turns out however
that if we introduce the  new basis

\begin{equation}\label{locope}
{\cal O}=q^{-\kappa-\frac{1}{2}}{\cal O}_{loc}\,,
\qquad
\tilde{{\cal O}}=q^{\kappa-\frac{1}{2}}\tilde{{\cal O}}_{loc}\,,
\qquad
q^2\equiv -2(p^+p^- + p\bar{p})
\end{equation}
then the representations of $psu(2,2|4)$ in ${\cal O}$
and $\tilde{{\cal O}}$ coincide. We found the following realization of
$psu(2,2|4)$ superalgebra in terms of differential operators acting on CFT
superfields ${\cal O}$, $\tilde{{\cal O}}$
\begin{equation}\label{cftgen1}
P^\pm =p^\pm,
\quad
P=p\,,
\quad
\bar{P}=\bar{p},
\quad
J^{+x}=\partial_p p^+ -\partial^+ p,
\quad
J^{+\bar x}=\bar{\partial}_p p^+ -\partial^+ \bar{p}\,,
\end{equation}
\begin{equation}\label{cftgen2}
K^+=K_0^+-\frac{3}{2}\partial_p^+
+\frac{p^+}{2q^2}A\,,
\end{equation}
\begin{equation}\label{cftgen3}
Q^{+i}=p^+\theta^i,
\quad
\quad Q^+_i=\lambda_i,
\quad
Q^{-i}=\bar{p}\theta^i+\frac{{\rm i}}{\sqrt{2}}q\chi^i,
\quad
Q^-_i=\frac{p}{p^+}\lambda_i-\frac{{\rm i}}{\sqrt{2}p^+}q\tau_i,
\end{equation}
\begin{equation}\label{cftgen4}
S^{+i}=\theta^ip^+\bar{\partial}_p
+{\rm i}\frac{p^+}{2\sqrt{2}q}[\chi^i,A]
-\partial^+_pQ^{-i}\,,
\qquad
S_i^+=-\lambda_i\partial_p+\frac{{\rm i}}{2\sqrt{2}q}[\tau_i,A]
+\partial_p^+Q^-_i\,,
\end{equation}
\begin{eqnarray}
&&
J^{+-}=-\partial_p^+p^- + \partial_p^-p^+-\frac{1}{2}\theta\lambda
-\frac{1}{2}\chi\tau+2\,,
\hskip6em
J^{x\bar{x}}=p\bar{\partial}_p-\bar{p}\partial_p
+\frac{1}{2}\theta\lambda-\frac{1}{2}\chi\tau\,,
\nonumber\\
\label{cftgen5}
&&
D=-\partial_p^+p^- - \partial_p^-p^+
-\partial_p\bar{p}-\bar{\partial}_pp
+\frac{1}{2}\theta\lambda+\frac{1}{2}\chi\tau
-\frac{1}{2}\,,
\qquad
J^i{}_j=l^i{}_j+M^i{}_j\,,
\end{eqnarray}
where the $AdS$ mass operator $A$ is given in (\ref{ax}), the $l^i{}_j$ and
$M^i{}_j$ are given in (\ref{lm}), while $K_0^+$ is given by
\begin{equation}\label{cftk0p}
K_0^+=-\partial_p\bar{\partial}_pp^+
-\partial^+_p(-\partial^+_pp^-
-\partial_p\bar{p}-\bar{\partial}_pp)\,.
\end{equation}
To be definite, here and below we assume that $q^2=p_0^2-\vec{p}^2>0$.
Expressions for $J^{-x}$, $J^{-\bar x}$ are obtainable from (\ref{jmx})
by inserting  there the expressions for $P^-$ and $Q^{-i}$, $Q^-_i$ given
in (\ref{cftgen1}), (\ref{cftgen3}). The remarkable property of
realization we constructed is that the dependence on $AdS$ mass
operator $A$ in CFT is `dual' to AdS representations.
Namely,  on $AdS$ side the $A$ appears in $P^-$ (\ref{adspm}) having
$-1$ $D$- and $J^{+-}$- charges, while on CFT side this operator appears
in $K^+$ (\ref{cftgen2}) having opposite, i.e., $+1$ $D$ and $J^{+-}$-
charges.  The same `duality' is the case of  AdS $Q^-$ generators
(\ref{adsqm}) having $-1/2$ $D$-and $J^{+-}$ charges
and CFT $S^+$ generators (\ref{cftgen4}) having opposite, i.e., $+1/2$
$D$- and $J^{+-}$-charges. As before,  the nonlinear dependence on
Grassmann variables $\chi^i$, $\tau_i$ is expressible through the operator
$A$. As was said already,   the  above representation is applicable  to
both  ${\cal O}$ and $\tilde{{\cal O}}$. The price for this is that the
generators are no longer local with respect to the  transverse momenta
$p$ and $\bar{p}$ included in $q$. However,  these nonlocal terms cancel
when we transform from ${\cal O}$ and $\tilde{{\cal O}}$ basis into the
one of ${\cal O}_{loc}$ and $\tilde{{\cal O}}_{loc}$.

\medskip
{\bf AdS/CFT correspondence}.
After we have derived the light cone formulation for both the bulk
superfield $\Phi$ and the boundary conformal theory operators collected in
${\cal O}$ and $\tilde{{\cal O}}$ we are ready to demonstrate explicitly
the AdS/CFT correspondence.  We demonstrate that boundary
values of normalizable solutions of bulk equations of motion are
related to conformal operators ${\cal O}_{loc}$, while the ones of
non-normalizable solutions are related to operators
$\tilde{{\cal O}}_{loc}$ (see \cite{Balasubramanian:1999sn},\cite{metlc}).
To this end let us consider solutions to light cone equations of motion which,
as usual, take the form $P^-\Phi=p^-\Phi$. Taking into account the expression
for $P^-$ (\ref{adspm}) and rewriting these equations as
$$
\Bigl(-\partial_z^2+\frac{1}{z^2}(\kappa^2-\frac{1}{4})\Bigr)\Phi
=q^2\Phi\,,
\qquad
\kappa\equiv \sqrt{X}\,,
$$
we immediately get the following normalizable and non-normalizable
solutions
\begin{equation}\label{solequmot}
\Phi_{norm}=\sqrt{qz}J_{\kappa}(qz){\rm i}^\kappa{\cal O}\,,
\qquad
\Phi_{non-norm}
=\sqrt{qz}\,Y_{\kappa}(qz){\rm i}^\kappa\tilde{{\cal O}}\,,
\end{equation}
where ${\cal O}$, $\tilde{{\cal O}}$ are scalar superfields do not
depending on $z$. The $J_\kappa$ and $Y_\kappa$ are Bessel and Neumann
functions respectively. The normalization factor ${\rm i}^\kappa$ is
included for convenience. In (\ref{solequmot}) we use the notation ${\cal
O}$, $\tilde{{\cal O}}$ since we are going to demonstrate that these
are indeed the CFT superfields discussed   above. Namely,  we
are going to prove that AdS transformations for $\Phi$ lead to conformal
theory transformations for ${\cal O}$ (and $\tilde{{\cal O}}$)
\begin{equation}\label{gadsgcft}
G_{ads}\Phi = Z_\kappa(qz){\rm i}^\kappa G_{cft}{\cal O}\,,
\qquad
Z_\kappa(z)\equiv \sqrt{z}J_\kappa(z)\,.
\end{equation}
Here and below we use the notation $G_{ads}$ and $G_{cft}$ to indicate
the realization of $psu(2,2|4)$ algebra generators on the  bulk field
(\ref{kin1})-(\ref{kin4}) and conformal operator
(\ref{cftgen1})-(\ref{cftgen5}) respectively.

Now let us make a comparison of generators for bulk field $\Phi$ and
boundary operator ${\cal O}$. Important technical simplification is that
it is sufficient to make comparison only for the part of the algebra spanned by
generators
$$ P^\pm,
\quad P,
\quad
\bar{P},
\quad J^{+-},
\quad J^{\pm x},
\quad J^{\pm \bar x},
\quad K^+,
\quad Q^{\pm i},
\quad Q^\pm_i\,,
\quad
J^{x\bar x},
\quad
D\,,
\quad
J^i{}_j\,.
$$
It is straightforward to see that if these generators match then the
remaining generators $K^-$ and $K$, $\bar{K}$, $S^{\pm i}$, $S^\pm_i$
shall match due to commutation relations of the $psu(2,2|4)$ superalgebra.
We start with a comparison of the kinematical generators (\ref{kingen}).
As for  the generators
$P^+$, $P$, $\bar{P}$, $J^{+x}$, $J^{+\bar x}$,
$J^{x\bar x}$, $J^i{}_j$, $Q^{+i}$, $Q^+_i$
they already coincide on both sides (see (\ref{kin1})-(\ref{kin4})
and (\ref{cftgen1})-(\ref{cftgen5})). In fact this
implies that the coordinates $p^+$, $p$, $\bar{p}$, $y^{ij}$, $\lambda_i$
and $\tau_i$ we use on AdS and CFT sides match.

Now let us consider $P_{ads}^-$ (\ref{adspm}) and $P_{cft}^-$
(\ref{cftgen1}). Taking into account that solutions to equation of motion
(\ref{solequmot}), by definition, satisfy the relation
$P_{ads}^-\Phi=p^-\Phi$ we get $P_{ads}^-\Phi=Z_\kappa(qz){\rm i}^\kappa
P_{cft}^- {\cal O}$.  So $P^-_{ads}$ and $P_{cft}^-$ also match. Taking
this into account it is straightforward to see that the generators
$J_{ads}^{+-}$, $D_{ads}$ (\ref{adsjpcom}),(\ref{adskmcom}) and
$J^{+-}_{cft}$, $D_{cft}$ (\ref{cftgen5})
satisfy the relation (\ref{gadsgcft}).
Next we consider kinematical generators $K^+_{ads}$ (\ref{adskmcom}) and
$K_{cft}^+$ (\ref{cftgen2}).  Here we use the following relation
$$
K_{ads}^+ Z_\kappa(qz)
=Z_\kappa(qz)\Bigl(K_{0cft}^+ +\frac{p^+}{2q^2}A
-\partial_p^+(\frac{3}{2}+z\partial_z)\Bigr)
+\frac{p^+}{q}(\partial_qZ_\kappa(qz))z\partial_z\,,
$$
where $K_{ads}^+$ and $K_{0cft}^+$ are given in (\ref{adskmcom}) and
(\ref{cftk0p}) respectively. Using then (\ref{solequmot}), the above
relation and the fact that ${\cal O}$ in (\ref{solequmot}) does not depend
$z$ we get immediately that $K_{ads}^+$
and $K_{cft}^+$ satisfy the relation (\ref{gadsgcft}).

The last step is to match the generators $Q^{-i}_{ads}$,
$Q^-_{i\, ads}$ and $Q^{-i}_{cft}$, $Q^-_{i\,cft}$.
This is the most challenging
part of the  analysis. Let us consider $Q^{-i}$. Generalization of
our discussion to the case of $Q^-_i$ is straightforward. As is seen from
(\ref{adsqm}) and (\ref{cftgen3}) requiring that these supercharges
satisfy the basic relation (\ref{gadsgcft}) gives the following nontrivial
relation
\begin{equation}\label{nontrirel}
(\chi^i\partial_z+\frac{1}{2z}[\chi^i,X])\sqrt{z}J_{\kappa}(z)
{\rm i}^\kappa
=\sqrt{z}J_{{\kappa}}(z){\rm i}^{\kappa} {\rm i}\chi^i\,,
\end{equation}
which is understood in weak sense, i.e., as an operator relation defined on
the space of superfield ${\cal O}$. This interesting relation is proved in
Appendix.  The same relation is valid in  the case of the Neumann function.
Taking into account matching $Q^{-i}$, $Q^-_i$ and $P^-$ we conclude that
$J^{-x}$, $J^{-\bar x}$ (\ref{jmx}) also match.  Above analysis is
obviously generalized to the case of non-normalizable solutions and shadow
operators (\ref{solequmot}).

Thus we have demonstrated that boundary operators in ${\cal O}$,
$\tilde{{\cal O}}$ in (\ref{solequmot}) are indeed the conformal
operators. Because of multiplicative factors $J_\kappa$ and $Y_\kappa$
involving powers of $q$ the operators ${\cal O}$ and $\tilde{{\cal O}}$,
however, are not boundary values of solutions of the  bulk equations of motion
(\ref{solequmot}). On other hand,  by relations (\ref{locope}) it is easily
seen that it is ${\cal O}_{loc}$ ($\tilde {\cal O}_{loc}$) that is the
boundary value of $\Phi_{norm}$ ($\Phi_{non-norm}$),  i.e.
for small $z$ one has the local interrelations
\begin{equation}\label{phiope}
\Phi_{norm}\sim z^{\kappa+\frac{1}{2}}{\cal O}_{loc}\,,
\qquad
\Phi_{non-norm}\sim z^{-\kappa+\frac{1}{2}}\tilde{{\cal O}}_{loc}\,.
\end{equation}
Note that for $\kappa=0$ the factor $z^{-\kappa}$ in the second relation of
(\ref{phiope}) should be replaced by $\log z$.

\medskip {\bf Conclusions}.
We have developed the light cone gauge formulation of $IIB$ supergravity
in $AdS_5\times S^5$ background and applied this formulation to the study
of AdS/CFT correspondence (for review see \cite{review}). Because the
formulation is given entirely in terms of the light cone scalar superfield it
allows us to treat {\it all}  fields of $IIB$ supergravity on equal footing and
in a manifestly  supersymmetric way.  Comparison of this formalism with other
approaches available in the literature leads us to the conclusion that
this is a very efficient formalism indeed.  The results presented here should
have a number of interesting generalizations, some of which are:

(i)
extension of light cone formulation of $IIB$ supergravity in
$AdS_5\times S^5$ background to the level of cubic interaction vertices
(see \cite{metpre}).

(ii) extension of light cone formulation of conformal field
theory to the level of OPE's and a study of light-cone form of AdS/CFT
correspondence at the level of correlation functions.

(iii) generalization of light cone gauge formalism to the study
of compactifications of 11-dimensional supergravity on $AdS_4\times S^7$
and $AdS_7\times S^4$ \cite{compact}. In these cases a formulation in
terms of light-cone scalar superfields could  also be  developed.

In view of  previous experiences with massless higher spin
fields in $AdS_4$ space it is known that to construct self-consistent
interactions for such fields it is necessary to introduce
%, among other things,
an infinite tower of  massless fields \cite{vas1}. For the case of
$AdS_5$ space it is expected that fields of $IIB$ supergravity constitute
lower spin multiplet of the infinite tower of $N=8$ supersymmetric
massless
higher spin fields theory in $AdS_5$ background.
In this perspective,  we think that the results
of this paper can also get interesting applications to massless higher spin
field theory. Note that $AdS_5\times S^5$ is a unique space where the
consistent string and massless higher spin field theories may `meet'.

\medskip
{\bf Acknowledgments}
I would like to thank A. Tseytlin for reading
the manuscript
 and making comments improving it.
This  work was supported in part
by INTAS grant No.96-538 and the Russian Foundation for Basic Research
Grant No.99-02-16207.

\medskip
{\bf Appendix}.
In order to prove (\ref{nontrirel}) we start with
$$
\chi_t^i\equiv e^{t\tau l\chi}\chi^ie^{-t\tau l\chi}\,,
\qquad
\tau_{i\,t}\equiv e^{t\tau l\chi}\tau_ie^{-t\tau l\chi}\,,
\qquad
l^i{}_{jt}\equiv e^{t\tau l\chi}l^i{}_je^{-t\tau l\chi}\,.
$$
Making use of the relations
$$
\partial_t \chi_t^i=-(l_t\chi_t)^i\,,
\qquad
\partial_t \tau_{i\,t}=(\tau_tl_t)_i\,,
\qquad
\partial_t l^i{}_{jt}=(\tau_tl_t)_j\chi_t^i-\tau_{jt}(l_t\chi_t)^i\,,
$$
where $(l\chi)^i\equiv l^i{}_j\chi^j$ (see also (\ref{somrel}),
we get a closed differential equation for $\chi_t^i$
which can be written in the following two equivalent forms
\begin{equation}\label{difequ1}
\partial_t^2\chi_t^i
=(\frac{1}{4}l^n{}_m^2+\tau l\chi)\chi_t^i
+(2-\chi\tau)(\partial_t\chi_t)^i\,,
\quad
\partial_t^2\chi_t^i
=\chi_t^i (\frac{1}{4}l^n{}_m^2+\tau l\chi)
+(\partial_t\chi_t)^i (2-\chi\tau)\,,
\end{equation}
where the initial conditions are $\chi_{t=0}^i=\chi^i$,
$\partial_t\chi_t^i|_{t=0}=-(l\chi)^i$.
Solutions to these equations are
\begin{eqnarray*}
&&
\chi_t^i=e^{\frac{t}{2}(2-\chi\tau)}
\Bigl(\cosh\frac{{\kappa}}{2}t\chi^i
+\frac{\sinh\frac{{\kappa}}{2}t}{{\kappa}}
((\chi\tau-2)\chi^i-2(l\chi)^i)\Bigr)\,,
\\
&&
\chi_t^i=\Bigl(\chi^i\cosh\frac{{\kappa}}{2}t
+(\chi^i(\chi\tau-2)-2(l\chi)^i)
\frac{\sinh\frac{{\kappa}}{2}t}{{\kappa}}\Bigr)
e^{\frac{t}{2}(2-\chi\tau)}\,.
\end{eqnarray*}
Equating these two different forms of the
solution we get the following basic
formula
$$
\chi^i\cosh\kappa t +(2(l\chi)^i+\chi^i(2-\chi\tau))
\frac{\sinh\kappa t}{\kappa}
=e^t\Bigl(\cosh\kappa t\chi^i
+\frac{\sinh \kappa t}{\kappa}
(2(l\chi)^i+(2-\chi\tau)\chi^i)\Bigr)\,.
$$
Inserting $t={\rm i}\varphi$ and taking real and imaginary parts gives
\begin{eqnarray}
\label{for1}
&&
{}\hskip-5ex
\chi^i\cos{\kappa}\varphi=\cos\varphi\cos{\kappa}\varphi\chi^i
+\frac{\sin\varphi\sin{\kappa}\varphi}{{\kappa}}
((\chi\tau-2)\chi^i-2(l\chi)^i)\,,
\\
\label{for2}
&&
{}\hskip-5ex
(2(l\chi)^i+\chi^i(2-\chi\tau))
\frac{\sin{\kappa}\varphi}{{\kappa}}
=\frac{\cos\varphi\sin{\kappa}\varphi}{{\kappa}}
(2(l\chi)^i+(2-\chi\tau)\chi^i)
+\sin\varphi\cos{\kappa}\varphi\chi^i\,.
\end{eqnarray}
Now we use the integral representation for Bessel function
$$
J_{\kappa}(z)=\frac{{\rm i}^{-{\kappa}}}{\pi}\int_0^\pi d\varphi\,
e^{{\rm i}z\cos\varphi}\cos{\kappa}\varphi\,,
$$
which is valid for integer $\kappa$. Taking into account the relation
$$
(\chi^i\partial_z+\frac{1}{2z}[\chi^i,X])\sqrt{z}J_{{\kappa}}(z)
=\sqrt{z}\Bigl(\chi^i\partial_z
+\frac{1}{z}(2(l\chi)^i+\chi^i(2-\chi\tau))\Bigr)J_{{\kappa}}(z)
$$
and the formulas (\ref{for1}), (\ref{for2}) we get
$$
\chi^i\partial_zJ_{{\kappa}}(z){\rm i}^\kappa
=\frac{1}{\pi}\int_0^\pi d\varphi e^{{\rm i}z\cos\varphi}
{\rm i}\cos\varphi\Bigl(
\cos\varphi\cos{\kappa}\varphi\chi^i
+\frac{\sin\varphi\sin {\kappa}\varphi}{{\kappa}}
((\chi\tau-2)\chi^i-2(l\chi)^i)\Bigr)\,,
$$
\begin{eqnarray*}
&&
\frac{1}{z}(2(l\chi)^i+\chi^i(2-\chi\tau))J_{{\kappa}}(z)
{\rm i}^\kappa
\\
&&
=\frac{1}{\pi}\int_0^\pi d\varphi e^{{\rm i}z\cos\varphi}
{\rm i}\sin\varphi\Bigl(
\sin\varphi\cos{\kappa}\varphi\chi^i
+\frac{\cos\varphi\sin{\kappa}\varphi}{{\kappa}}
((2-\chi\tau)\chi^i+2(l\chi)^i)\Bigr)\,,
\end{eqnarray*}
To derive the last formula we integrate by parts and since
the operator $\kappa$ takes integer values the boundary terms cancel.
Summing these contributions we get the desired relation (\ref{nontrirel}).

\end{document}